\documentclass[10pt]{bmc_article}

\usepackage{cite} 
\usepackage{url}  
\usepackage{ifthen}  
\usepackage{multicol}   
\usepackage[utf8]{inputenc} 
\usepackage{graphicx}

\newboolean{publ}

\newenvironment{bmcformat}{\begin{raggedright}\sloppy\setboolean{publ}{false}}{\end{raggedright}\baselineskip20pt\sloppy}

\begin{document}
\begin{bmcformat}


\title{Genetic flow directionality and geographical  segregation in a
{\normalfont\bfseries\itshape Cymodocea nodosa} genetic  diversity network}


\author{Paolo Masucci$^1$%
         \email{Paolo Masucci - a.masucci@ucl.ac.uk}%
       \and
       Sophie Arnaud-Haond$^2$%
        \email{Sophie Arnaud-Haond - sophie.arnaud@ifremer.fr}%
       \and
       V\'{\i}ctor M. Egu\'{\i}luz$^3$%
        \email{V\'{\i}ctor M. Egu\'{\i}luz - victor@ifisc.uib-csic.es}%
       \and
       Emilio Hern\'{a}ndez-Garc\'{\i}a\correspondingauthor$^3$%
         \email{Emilio Hern\'andez-Garc\'{\i}a\correspondingauthor - emilio@ifisc.uib-csic.es}%
       and
       Ester A. Serr\~{a}o$^4$
        \email{Ester A. Serr\~{a}o - eserrao@ualg.pt}
        }


\address{%
    \iid(1) Centre for Advanced Spatial Analysis, University College of London, London, UK\\
    \iid(2) Institut Fran\c{c}ais de Recherche pour l'Exploitation de la MER, D\'{e}partement \'{E}tude des
    Ecosyst\`{e}mes Profonds-DEEP, Laboratoire Environnement Profond-LEP, Centre de Brest, France\\
    \iid(3) IFISC (CSIC-UIB), Instituto de F\'{\i}sica Interdisciplinar y Sistemas Complejos,
    Consejo Superior de Investigaciones Cient\'{\i}ficas - Universitat de les Illes Balears,
    Palma de Mallorca, Spain\\
    \iid(4) CCMAR, CIMAR-Laborat\'{o}rio Associado, Universidade do Algarve, Gambelas, 8005-139, Faro, Portugal
}%

\maketitle


\begin{abstract}
We analyse a large data set of genetic markers obtained from
populations of \textsl{Cymodocea nodosa}, a marine plant
occurring from the East Mediterranean to the Iberian-African
coasts in the Atlantic Ocean. We fully develop and test a
recently introduced methodology to infer the directionality of
gene flow based on the concept of geographical segregation.
Using the Jensen-Shannon divergence, we are able to extract a
directed network of gene flow describing the evolutionary
patterns of \textsl{Cymodocea nodosa}. In particular we recover
the genetic segregation that the marine plant underwent during
its evolution. The results are confirmed by natural evidence
and are consistent with an independent cross analysis.
\end{abstract}

\ifthenelse{\boolean{publ}}{\begin{multicols}{2}}{}


\section*{Introduction}
With the advances of sequencing technology and the availability
of large datasets, evolutionary biology needs to employ novel
techniques, which are akin to those developed within
statistical physics \cite{drossel2001}, to analyse and
understand patterns in population dynamics. An interesting
question in evolutionary biology is how to trace the
directionality in migration patterns, a problem of outstanding
importance in conservation biology in general, including the
management of threatened or exploited species and of invasion
processes \cite{Travis2004}. It is also related to the more
general problems of infection and information propagation in
networks \cite{Haydon2002,Vespignani2012}.

In this paper we tackle the indirect assessment of migratory
transfers (by pollen, propagules, or plant fragments) among
plant populations using molecular markers to retrace the
exchange of genes, or {\sl gene flow}, among them. This is done
by fully developing a recently introduced methodology based on
a directionality index \cite{masucci}. This index finds its
origins on the concept of \textit{geographical segregation}
\cite{Duncan55}, often related to social studies, to infer the
evolutionary pathways from microsatellite datasets.
Microsatellites \cite{messier} are portions of non-coding DNA
with a variable number of repetitions of a motif consisting of
a few bases. They are widely used in intraspecific genetic
studies. Despite being non-coding, we will use eventually the
word {\sl genes} to denote the microsatellites, and {\sl
alleles} for their different varieties occurring at a
particular position in the genome (a particular {\sl locus}).


Classical population genetics analyses do not allow inferring
the direction of migration using molecular data. Although some
Bayesian analyses have been more recently developed to do so
\cite{Beerli1999,Guillemaud2010}, many require complex and
time-consuming computing of the likelihood functions,
restraining the ability to explore more than often too simple
evolutionary scenarios and molecular models
\cite{Csillery2010}.

In the present study we analyse a dataset that presents evident
cases of geographical segregation, such as island effects and
we are able to show that the proposed methodology spots these
islands, based on microsatellite data only and without any
further geographical information or evolutionary assumption. In
particular we use the information contained in microsatellite
genetic markers from the entire geographic distribution of a
marine plant species, \textsl{Cymodocea nodosa} (CN). This
dataset has been selected because there is enough information
to infer the past history of the gene flow based on the
geographical distribution of genetic polymorphism
\cite{alberto2008}, allowing the assessment of the usefulness
of the new methods here described.

Understanding the pathways of gene flow along the
Mediterranean-Atlantic transition zone was the main aim of this
genetic dataset \cite{alberto2008}. Based on presence/absence
of alleles, this revealed that the flow of genetic information
across the Mediterranean-Atlantic transition zone had most
likely occurred westwards, because dominant Mediterranean
alleles penetrate into the nearest Atlantic sites (Atlantic
Iberia), but the opposite is not true, i.e. dominant Atlantic
alleles are not found in any Mediterranean populations.  This
clear pattern of presence/absence of diagnostic alleles results
in that this data set provides an ideal workbench to test and
to develop a recently introduced method of inferring
directionality of gene flow, here based on distances computed
from the Jensen-Shannon divergence \cite{lin1991}.

Network theory has already proved to be useful in the study of
metapopulation systems dynamics \cite{rozenfeld}. In particular
it has been shown that the analysis of topological
relationships between different populations carries fundamental
information for the understanding of evolutionary dynamics
\cite{rozenfeld}. It is important to further develop and test
methodologies to extract reliable information-flow networks
from biological datasets.

The method considered in this work introduces some novelties
with respect to classical approaches, that have been already
underlined in \cite{masucci}. The gene flow network is
extracted by means of a model-independent measure, that is a
normalised version of the Jensen-Shannon divergence. An
interesting aspect of the application of this method is that we
extract the \textit{gamete space} from the allele sequences and
we perform the analysis in that space. {\sl Gametes} are the
mother and father sexual cells that fuse in a sexual
reproduction event. The nuclear genetic information from the
two gametes remain mostly separated in the daughter cell (in
the different chromosomes of each chromosome pair) and
subsequent cells originating from it, but sequencing techniques
can not identify which gene is coming from which gamete. From
the observed alleles in each individual in our dataset we
construct the set of possible gametes that could have
originated such individual, and we apply our techniques to such
gamete pool. The distance measure in the gamete space has to be
considered more detailed than the one in the allelic space,
since it takes into account the possible correlations between
different loci. An extension of the method to include mutation
effects can be easily obtained as better explained below.

Here we advance the method beyond its first application
\cite{masucci}, in fully developing and verifying the
directionality index methodology by introducing a test for the
detected directionality significance. This is done with an
ad-hoc randomization test. This method is independent of the
way the genetic flow is extracted and can then be applied
independently of the genetic distance used. The present
analysis reveals that the methodology is  efficient in cases of
evident geographical segregation, for which the method was
designed, while it is not efficient in detecting the direction
of the flows where geographical segregation is not present.

Moreover in the {\sl Additional file 1}, we give a detailed
comparison of the distance method applied here with some of the
most used genetic distance measures based on microsatellite
analysis, such as the Nei distance  \cite{3}, the
Cavalli-Sforza distance \cite{2}, the Goldstein distance
\cite{3} and the average square distance \cite{4}.

\section*{Data and results}
\subsection*{The genetic flow network}
\label{gfc}

\begin{figure}[!ht]\center
\includegraphics[width=.9\textwidth]{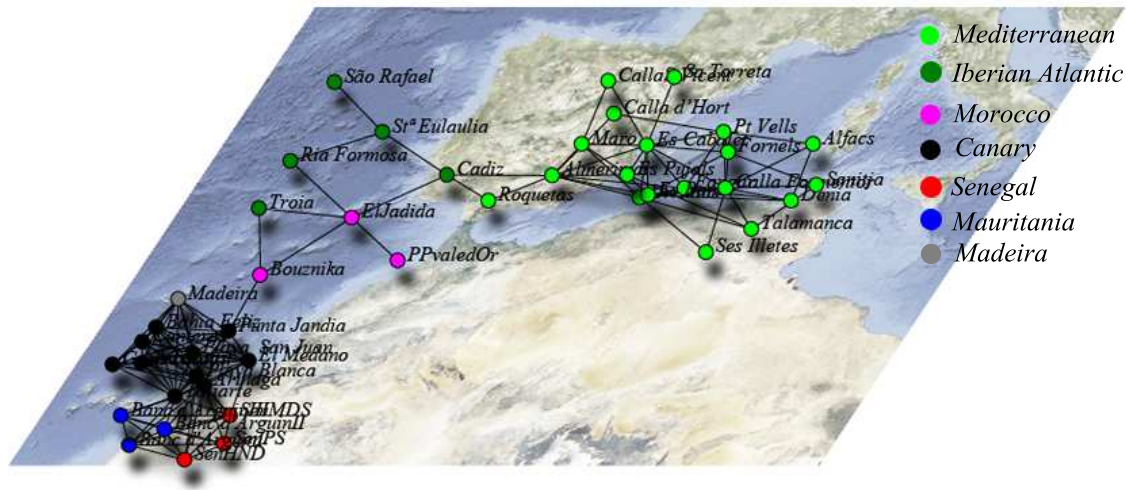}
\caption{{\bf Figure 1 - Genetic flow network for
      {\normalfont\bfseries\itshape Cymodocea nodosa}.}
      Genetic flow network for \textsl{Cymodocea nodosa}
      meadows at the percolation threshold, obtained via
      the JSD measure applied to raw gamete data from the
      sampled meadows (the genetically very distant and
      disconnected Greek populations are not shown).
      Different colors indicate the different meadow
      origins. The network is displayed via a spring
      embedding algorithm, i.e. it does not contain
      geographical information. Nevertheless the genetic
      clusters well trace the different geographical
      origins, as it is possible to see by comparison with
      the approximately overlayed map displayed below.}
\end{figure}

Our dataset consists of 845 \textit{ramets} of CN. A ramet is a
single plant shoot, whereas a {\sl genet} is a genetic
individual, or clone, derived from a single event of sexual
recombination (i.e., from a single seed) and having given birth
through clonal growth to a population of ramets therefore
sharing the same genome \cite{Arnaud2007}. Ramets were sampled
from 60 different meadows, distributed geographically between
the Mediterranean basin and the Atlantic Ocean, covering the
entire plant distribution \cite{alberto2008}. This dataset is
the one discussed in \cite{alberto2008} with the addition of a
few populations more recently sampled, as for example from
Morocco. Among the 40 ramets taken at each site, after removal
of genet (i.e., clonal) repetitions the number of ramets
available in the data set ranged from 4 to 34 per meadow. Hence
each of the 845 ramets of the dataset represents a different
clone.

We characterize each ramet by some microsatellite markers
\cite{messier}. In particular, each ramet has been genotyped to
identify in it $n=8$ pairs of alleles, i.e.  pairs of
microsatellites that occupy a specific position (locus) on the
chromosomes, each element of the pair characterizing the same
locus in the two homologous chromosomes arising from the
maternal and paternal gametes in a sexual reproduction event.
The number $n$ of pairs of microsatellites used was selected
for highest information content with minimum cost
(\cite{alberto2008} and references therein), a standard
microsatellite genotyping methodology.

To characterize the presence of gene flow between meadows we
use the general network strategy
\cite{Dyer2004,rozenfeld,DaleFortin2010}, in which populations
(here, the meadows) are nodes of a graph, and they are linked
when significant relationships among them (indicating gene
flow) are detected.

There are different ways to implement such inference of gene
flow, mostly based on different types of distances
\cite{Dyer2004,rozenfeld,Fortuna2009}. Here we  use a
methodology based on information theory \cite{masucci}, which
is especially suited to compare genetic data taken from
populations of different sizes, taking into account not just
properties of individual alleles but also the full information
genotyped (including correlations among sites, linkage
disequilibrium, for example).

We consider the set of meadows as a metapopulation system where
each population is a meadow and each population element is a
\textit{n}-dimensional vector representing an equiprobable
gamete of that meadow, where \textit{n} is the number of loci.

To derive the gamete pool we notice that each ramet is
characterised by a set of \textit{n=8} pairs \textit{(a,b)} of
alleles, each pair belonging to a given locus. Since it is not
known \textit{a-priori} which chromosome a given allele belongs
to, we consider all the possible combinations of alleles
between the \textit{n} pairs, i.e. $2^n=256$ equiprobable
gametes representing a single ramet. In this way our system
becomes represented by $845\cdot 2^8=216320$ points in a
\textit{8}-dimensional space. Each meadow can be characterised
by a discrete probability function $P=P(\overrightarrow x)$
assigning a relative weight to each of the possible gametes
$\overrightarrow x$ in \textit{n}-dimensional space, i.e.
$P=P(\overrightarrow x)$ is the probability to find the gamete
$\overrightarrow{x}$ in that meadow.

For each pair of meadows characterised by probability
distributions $P=P(\overrightarrow x)$ and $Q=Q(\overrightarrow
x)$ we calculate the normalised Jensen-Shannon divergence
\begin{equation}\label{eq1}
JSD(P\|Q)\equiv \frac{H(\pi_AP+\pi_BQ)-\pi_AH(P)-\pi_BH(Q)}{ -\pi_A \ln \pi_A-\pi_B \ln \pi_B},
\end{equation}
With $\pi_{A,B}=n_{A,B}/(n_A+ n_B)$, $n_i$ the sample size of
meadow \textit{i} and $H(F)=-\sum_{\overrightarrow x}
F(\overrightarrow x) \ln F(\overrightarrow x)$ the Shannon
index or entropy of distribution $F$.

The information-theoretic meaning of \textit{JSD} is discussed
in detail in \cite{lin1991,Stanley2002, masucci}. We stress
that \textit{JSD} is a measure of difference between $P$ and
$Q$ that takes into account the information on the gametes
which are shared by both populations as well as the ones which
are exclusive to one of them. This second capability is not
present in other information-theoretic distances, such as the
Kullback-Leibler divergence \cite{lin1991}.

In the {\sl Additional file 1}, we illustrate in detail the
relationships between Eq. (\ref{eq1}) and other classical
measures to calculate genetic divergence. We want to stress
that Eq. (\ref{eq1}) is independent of any evolutionary
assumption, i.e. it just calculates the punctual correlations
within the meadows' attribute probability distributions.
Nevertheless it is possible to relax and extend such a measure
to include mutation effects, just considering the computation
of the probability distribution in a \textit{n}-dimensional
sphere of a given radius, surrounding the point
$\overrightarrow{x}$. Varying such a radius, it is then
possible to coarse-grain the system at different resolutions.

\textit{JSD} distances are calculated among all pairs of
meadows. Smaller distance implies stronger genetic identity
among the meadows. By selecting a particular threshold value
for the distance we can represent the genetic flow as a network
\cite{rozenfeld, masucci} in which meadows with smaller
\textit{JSD} distance appear linked. As we increase the
threshold we observe how the different linked clusters of
populations become larger and merge.

A convenient threshold to use is the percolation threshold
\cite{rozenfeld}, at which a connected path across the whole
geographic area first appears. The network fragments for
threshold values below the percolation threshold, while the
genetic flow spanning the network remains robust above it. At
this point different clusters, and subclusters, representing
sets of meadows with important internal gene flow, and gene
paths among them, can be identified.

In Fig. 1 we show the network at the percolation threshold,
when the major components in the network connect. As we can see
the purely topological genetic flow network accurately reflects
the geographical locations to which the different meadows
belong. There are well connected clusters representing Senegal,
Mauritania, Canaries and Madeira, then we have the other large
cluster spanning within the Mediterranean meadows. Between
those two big clusters we find the Atlantic Iberian meadows and
the Moroccan meadows.

\begin{figure}[!ht]\center
\includegraphics[width=.7\textwidth]{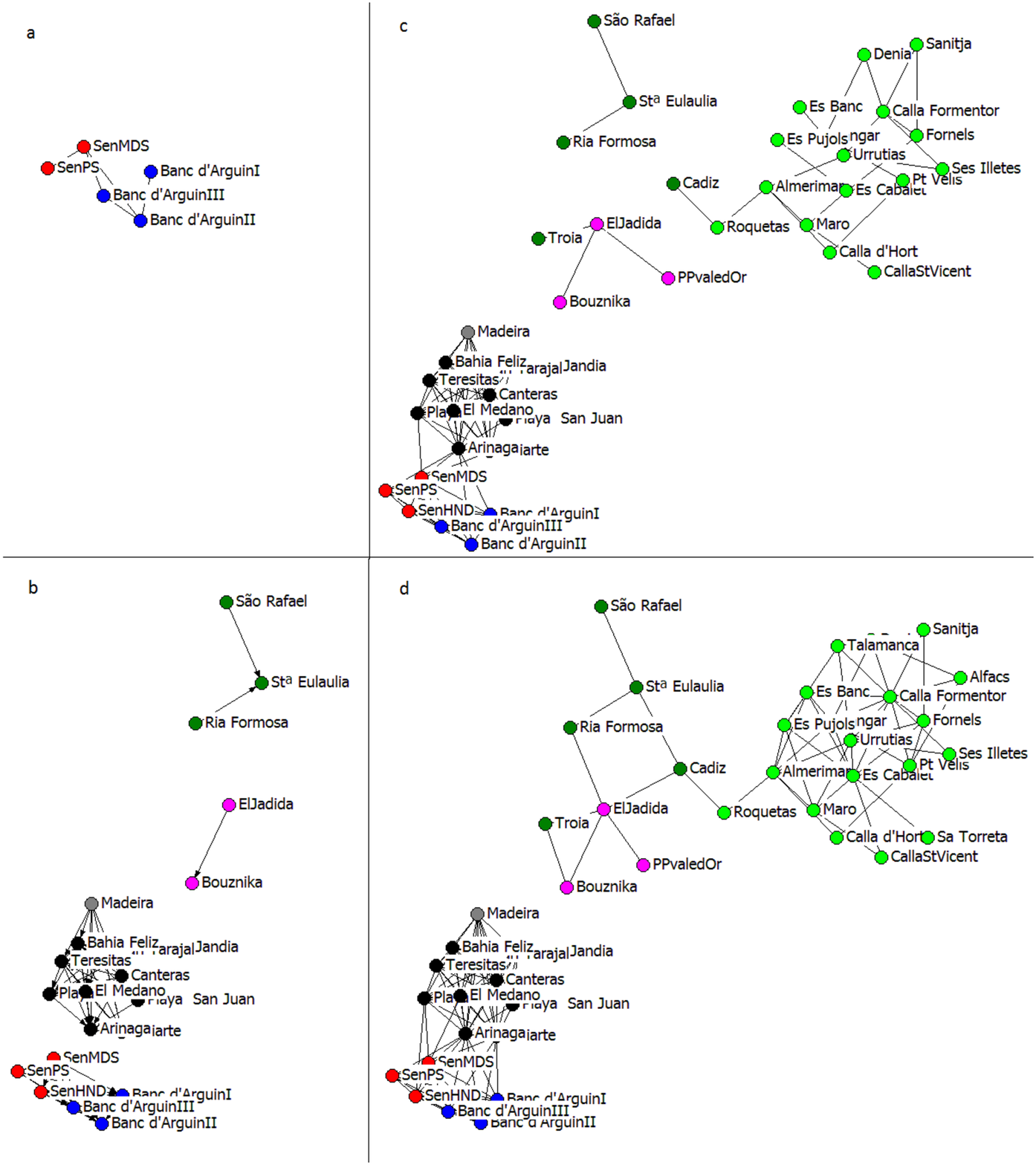}
\caption{{\bf Figure 2 - Genetic flow network for
    {\normalfont\bfseries\itshape Cymodocea nodosa} at
    growing values of the distance threshold.} Clusters
    that appear when linking populations with a JSD
    distance smaller than a distance threshold $T$, for
    increasing values of $T$ approaching from below the
    percolation threshold. a) $T=0.57$. b) $T=0.89$. c)
    $T=0.95$. d) $T=0.97$, just below the percolation
    threshold. The cluster colors are the same as in Fig.1.}
\end{figure}

While this representation is interesting to understand the
effectiveness of \textit{JSD} as a genetic divergence measure,
it also confirms the findings of \cite{alberto2008} regarding
the main evolutionary scenario for CN. To see that we plot in
Fig. 2 the connected clusters that form when increasing the
distance threshold. We interpret that the first populations to
merge when the genetic distance threshold is raised from zero
are the ones that differentiated more recently. Then small
genetic divergences correspond to the most recent time of
divergence. In our analysis we observe that the first meadows
to be connected are the ones South of the Canaries, while the
ones that are more distant are the ones in Greece (those do not
appear in the figure, since they are not connected until higher
distance thresholds), with a range of intermediate distances in
between. Then we can infer an evolutionary dynamics that starts
in the Greek Sea, goes to the main Mediterranean basin and then
spreads in the Atlantic. This coincides with the ancient
history of habitat colonization by this species inferred in
\cite{alberto2008}, which proposes that the species originated
in the eastern Mediterranean by divergence from its close
relative in the Indian Ocean/Red Sea and colonized the western
Mediterranean and Atlantic by spreading westwards. Moreover the
fact that Senegal and Mauritania cluster strongly with the
Canaries and Madeira, in respect to the weak clustering of the
Canaries with Morocco and the Iberian Atlantic, is in agreement
with the evolutionary scenario previously suggested on the
basis of biogeographical information. These findings are also
in agreement with the possible extinction of meadows in Morocco
and Iberia during the last glacial maximum, that was
accompanied by a drop in sea surface temperatures below the
range at which CN commonly occurs, and by a drop of sea level
that changed coastline morphology \cite{Thiede1978}, that would
have been followed by colonization of this region by an
admixture of Atlantic and Mediterranean genetic types
\cite{alberto2008}.

\subsection*{Inferring genetic flow directionality}

The second part of the analysis is about inferring the
directionality of the detected gene flows. The main question we
pose is: ``is it possible to understand which is the source and
which the sink in a given genetic channel?". To understand
this, we further develop the technique introduced in
\cite{masucci} based on geographical segregation
\cite{Duncan55}. We say that a population is segregated when it
contains elements (gametes in our case, but this could equally
be applied to alleles, as shown in the {\sl Additional file 1})
quite exclusive and distinct from the rest of populations, and
elements common in other populations are not so abundant here.
The main reasoning resides on the observation that a population
which is initially segregated will not maintain its character
if it is open to receive gametes from other different ones.  It
will remain segregated only if there is no gene exchange or if
there is some but the population acts as a source.

\begin{figure}[!ht]\center
\includegraphics[width=.9\textwidth]{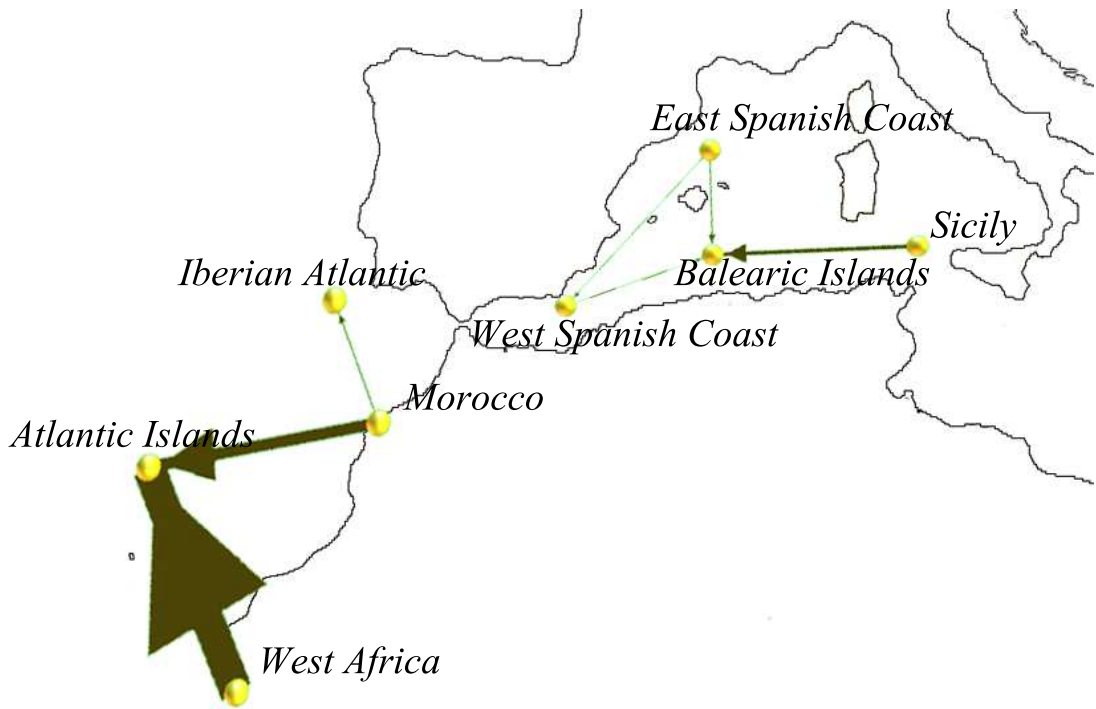}
\caption{{\bf Figure 3 - Gene flow directionality network. }
    Gene flow directionality network for \textsl{Cymodocea
    nodosa} genetic flow network evaluated in the gamete
    space. Nodes represent populations of meadows grouped
    together by their geographical attributes. Arrows are
    directional genetic flows, their width is proportional
    to the value of R, quantifying the significance of the
    inferred directionality.}
\end{figure}

In terms of frequency distributions, these are peaked on
particular elements in segregated populations, whereas
distributions are more unstructured in those acting as sinks
receiving genes from different sources. Between any pair of
populations among which a genetic flow has been detected (as
described in the previous section, this occurs when their
genetic distance is smaller than the given percolation
threshold), a directionality index $ I_{PQ} $ is defined
between meadows characterized by gamete distributions
\textit{P} and \textit{Q}. It takes into account the different
segregation state of the pairs of populations by means of the
respective Shannon information indices, corrected for their
different sizes and number of common elements.

To be more precise, given the two populations characterized by
$P$ and $Q$ let us denote by $D_P$ the set of different gametes
present in the first population, $D_Q$ the ones present in the
second, $X=D_P\bigcup D_Q$ the total set between both
populations, and $J=D_P\bigcap D_Q$ the set of common gametes
(that we assume to be non-empty). We denote by $\mu_P$ the
fraction between the number of gametes common to both
populations with respect to the total number of gametes in the
first one, and analogously we define $\mu_Q$ as the fraction of
common gametes between both populations with respect to the
total number of gametes in the second one. We have
$0<\mu_{P,Q}\leq 1$ and if $\mu_P$ or $\mu_Q$ is close to one,
it means that the shared alleles are the dominant part of the
corresponding population. Let us denote by
$P_J=P_J(\overrightarrow x)$, with $\overrightarrow x \in J$,
the frequency distribution of the alleles in the first
population, normalized to unity when summing over all the
common alleles (the $J$ set), and analogously we define
$Q_J=Q_J(\overrightarrow x)$, with $\overrightarrow x \in J$.

With these definitions, the directionality index is defined as
\cite{masucci}:
\begin{equation}
I_{PQ}\equiv -{\rm sign} \left[\frac{H(P_J )}{\mu_P}-\frac{H(Q_J )}{\mu_Q} \right]
\end{equation}
where \textit{H} is the Shannon index of the distributions, as
defined in the previous subsection.

If $I_{PQ}=1$ we can infer a direction of the genetic flow from
the first to the second population, whereas $I_{PQ}=-1$
indicates the reverse flow. The idea is that the net flow of
gametes is from the less entropic (i.e., less diverse, more
segregated) populations towards the more entropic, properly
adjusting for the different sizes and fraction of common
alleles. Very diverse populations can not be sources of only
very specific gametes, and they are more likely to be sinks,
whereas the reverse will be true for segregated populations.
Large width of gamete distributions may be an indicator of
large diversity, but it is better to use the Shannon
information as a more robust indicator of diversity. For more
information about our directionality indicator we refer to
\cite{masucci}.

\subsubsection*{Significance test of the detected directions}

The idea of tracing the directionality of gene flow via a
measure of geographical segregation turns out to be a delicate
point. First of all this method is applicable in case there are
traces of segregation in the subset of shared gametes of the
considered meadow. After that we have to understand if the
difference in the segregation indexes between the two meadows
in consideration is large enough to let us infer a direction
for the genetic flow.

To address these points first we should remember that the data
set used to infer directionality is not the whole set of
gametes, as used for network construction, but just those
gametes that are shared by the two samples in consideration. A
convenient way to quantify the number of shared gametes is by
means of the proportion $ \mu_P $ of shared gametes between
populations characterized by distributions \textit{P} and
\textit{Q}, with respect to the total number of gametes in the
population characterized by \textit{P}. We emphasize that $
\mu_P $ is not only a property of population \textit{P}, but of
the pair of populations being compared. This proportion should
be non-negligible in order to infer reliable conclusions for
the flow directionality.


Second, to understand if the detected directionality  $ I_{PQ}
$ between two meadows of size $n_1$ and $n_2$ is a sign of
segregation or it is a random effect, we first measure the
magnitude for the directionality index  $ I_{PQ} $ and then we
measure it again after having shuffled the sample. To shuffle
the sample we consider two meadows of size $n_1$ and  $n_2$ and
we fill them randomly with the genets extracted from the
original two meadows. We repeat this operation 1000 times, so
to obtain an estimation of the distribution for the randomised
directionality index, which turns out to be Gaussian. From
there one gets the average randomised directionality index
$I_{PQ}^R$ and its standard deviation $\sigma_R$.

The ratio $ R=|I_{PQ}-I_{PQ}^R |/\sigma_R $ gives us the number
of standard deviations (sd) the measured segregation index is
far from the expected one in a random situation. For instance
obtaining a value of \textit{R} larger than 2.58 has a
probability $p<0.01$ of being explained by randomness.

\begin{table}[htbp]
\begin{center}
\begin{tabular}{|c|r|r|r|}
\multicolumn{4}{c}{Randomisation test}\\\hline
\textbf{Link P-Q}&\textbf{\textit{R}}&\textbf{$\mu_P$}&\textbf{$\mu_Q$}\\\hline
\emph{\textbf{WA-AI}}&442.48&0.26&0.01\\
\emph{\textbf{MO-AI}}&159.16&0.03&0.004\\
\emph{\textbf{SI-BI}}&65.45&0.02&0.008\\
\emph{\textbf{ESC-BI}}&15.64&0.17&0.06\\
\emph{\textbf{SI-BI}}&65.45&0.02&0.008\\
\emph{\textbf{MO-IA}}&13.61&0.07&0.03\\
\emph{\textbf{ESC-WSC}}&4.92&0.09&0.07\\
\emph{\textbf{WSC-BI}}&4.4&0.24&0.13\\\hline
\end{tabular}
\end{center}
\caption{Results of the randomisation test for the
directionality analysis. Only the significant ($p<0.01$)
directionalities are listed. In the first column the population
pairs, in the second column the value of $R$ calculated over
1000 replicas, in the third column $\mu_P$, in the fourth
column $\mu_Q$. The acronyms stand for: WA: West African coast,
AI: Atlantic islands, MO: Morocco, SI: Sicily, BI: Balearic
Islands, ESC: East Spanish coast, IA: Iberian Atlantic, WSC:
West Spanish coast. } \label{t1}
\end{table}

\subsection*{Results}
To have a better statistical sample, we group the elements of
the clusters shown in Fig. 1 into single populations. Since we
are interested in retrieving the directionalities of gene flow,
we cluster the meadows according to their geographical origin.
Nevertheless such a clustering is not significatively different
from the one that would be obtained via a technique based, for
example, on modularity minimisation \cite{newmanmod}.

In Fig.3 we show the  results for the directional analysis. The
widths of the arrows are proportional to $R$, quantifying the
significance of the relationship. The genetic flow
directionality traces are found to be significant in terms of
$R$ and $\mu$ (see Tab. 1) from the Spanish Coast and Sicily to
the Balearic Islands, from the African coast to the Canary
Islands and Madeira, from the East (Mediterranean) Spanish
Coast to the West (Mediterranean) Spanish Coast and from
Morocco to the Iberian Atlantic. These results agree with
expectations from inferences based on allele presence/absence
by \cite{alberto2008} and \cite{alberto2006}.

These results tell us that the method correctly identifies
phenomena of geographical segregation. This phenomenon is
typical for islands, where it is not easy for species to
genetically mix with far away populations. Then we can see that
all the islands in the studied sample are identified as sinks
of the genetic flow. Also the main east toward west spread
trend has been identified by this methodology.

In contrast with the recent events of gene flow, this method is
not efficacious to trace the directionality from the
Mediterranean basin to the Atlantic, a much more ancient
process that took place before all the glacial range shifts and
post-glacial recolonization migrations, since there is no trace
of segregation in that case. Directionality of the genetic
flow, at such ancient scales, cannot be spotted with this
technique.

\section*{Discussion}

The directional network resulting from this  analysis is in
agreement with diverse and converging information on historical
fluctuation of species ranges along the Mediterranean-Atlantic
transition zone, associated to paleoclimatic events, putative
recolonization pathways and secondary contact zones inferred
from biogeographic analysis on various taxa, and expectation
derived from oceanographic modeling.

The directed network (Fig. 3) reveals a clear trend of gene
flow of South-Western preferential migration pathways from the
Western and central Mediterranean meadows towards the
Almeria-Oran front (an oceanographic feature east of the
Gibraltar strait). This coincides with the gene flow paths that
were previously hypothesized on the basis of many private
alleles found in the Atlantic that do not enter this transition
zone, whereas Mediterranean polymorphism was much more shared
with populations from the transition zones \cite{alberto2008}.
Modeled Lagrangian dispersal trajectories across the East-West
Mediterranean divide \cite{pinardi}) from March to June, the
fruit dispersal season for \textsl{Posidonia oceanica} (which
flowers in the winter), support a main trend of particle
exchange toward West in the Mediterranean \cite{serra}. This
was associated to a lack of dispersal expected toward Sicily
and the Eastern part of the Mediterranean, as predicted from
the prevailing current directions. Despite having a different
reproductive season (flowering in spring, seeds produced in
summer) and seeds that develop buried in the sediment and are
not expected to disperse with currents, CN is likely to
disperse via reproductive plant fragments, drifting along these
same current directions, in agreement with the inferences of
the present directionality flow network analysis, from Sicily
westwards.

Besides this direction of the Mediterranean flow toward the
Almeria-Oran populations, a second major result with the
presented methodology is the confirmed lack of input of the
Atlantic into the Mediterranean. Preferential pathways of
migration within the Atlantic show strong directionality of
flow from Western Africa and Morocco toward the Canary islands,
and from Morocco toward Iberian Atlantic populations of the
transition zone, with no sign of any entrance of Atlantic flow
into the Mediterranean.

In summary, the directed network built here confirms a dominant
direction of fluxes with a East-West direction in the
Mediterranean from Sicily and Spain toward South Western
meadows and Almeria-Oran Front, and the lack of mirror exchange
from the Atlantic toward the Mediterranean.

A phenomenon that can confound some of our techniques is that
of a genetic bottleneck. This consists of a large reduction of
population size, which leads also to a much decreased diversity
because of genetic drift\cite{gendrift}. This happens for
example if only a few individuals colonize a new island
(founder effect) and expand there. The recovered new population
has small diversity, with the potential to be identified as a
``source" although in fact it has been the receiver of the gene
flow. Fortunately this is not a problem because our methodology
does not compute the directionality index for all pairs of
populations, but only among those whith a strong similarity, as
measured by the JSD distance. Our measure of genetic flow will
give a very small value in the case mentioned above. In
addition, even if a recent and strong bottleneck (i.e.
contemporary to the sampling) or a recent re-expansion
post-bottleneck may transitorily induce ``source-like"
characteristics in the distributions, such distributions would
display a characteristic lack of polymorphism in the first
case, and a typical genetic signature on the distribution of
polymorphism and divergence in the latter that will be easily
detected by available population genetics tools
\cite{Luikart1998con,Luikart1998jher}. A careful examination of
data should thus allow discarding such confounding factor or
pointing it as a possible alternative explanation of
directionally detected with our methods.

We stress that the data treatment presented in this paper is
independent of evolutionary assumptions, but that it can be
easily extended to include mutation. This can be done, as we
already point out, coarse-graining the probability space in
function of a simple parameter indicating the resolution with
which we distinguish different gametes. Then varying this
parameter it is possible to study the evolutionary paths at
different time windows, but further research in this direction
is needed to better implement these points.

While the present research is about genetic flow dynamics, the
whole information flow/directionality index method has a wider
range of application, from metapopulation systems to social
dynamics and more in general it can be relevant whenever a
given population can be represented by vectors of attributes in
a symbolic space.

\bigskip


\section*{Competing interests}
The authors declare that they have no competing interests.

\section*{Author's contributions}
PM developed the methodology and performed the analyses. SAH
and EAS provided the data and the biological interpretations.
VME and EHG contributed to the methods and to the analyses. All
authors participated in the writing of the paper.

\section*{Acknowledgements}
  \ifthenelse{\boolean{publ}}{\small}{}
We thank Filipe Alberto for supplying the data, and discussing
the results. Supported by Ministerio de Econom\'{\i}a y
Competitividad (Spain) and Fondo Europeo de Desarrollo Regional
through projects FISICOS (FIS2007-60327) and MODASS
(FIS2011-24785). Publication fee has been covered partially by
the CSIC Open Access Publication Support Initiative through its
Unit of Information Resources for Research (URICI).


\newpage
{\ifthenelse{\boolean{publ}}{\footnotesize}{\small}
 \bibliographystyle{bmc_article}  
  \bibliography{cym} }     


\ifthenelse{\boolean{publ}}{\end{multicols}}{}



%



\section*{Additional file 1}

Here we compare $JSD$ as defined in Eq.1 of the main text with
some of the most used genetic distances. In particular we
compare it with the Nei distance $NEI$ \cite{2}, the
Cavalli-Sforza distance $CS$ \cite{1}, the Goldstein distance
$GD$ \cite{4} and the average square distance $ASD$ \cite{3}.
$D$, $NEI$ and $CS$ are distances defined in a symbolic space,
while $GD$ and $ASD$ are defined in metrical space where the
metric is defined by the allele repetitions.

The \textbf{Cavalli-Sforza distance} $CS$ is defined as
\begin{equation}\label{cs}
CS=\sqrt{4\frac{\sum_l(1-\sum_i\sqrt{x_iy_i})}{\sum_l(a_m-1)}}.
\end{equation}
Here $x_i$ is the fraction of allele $i$ in the first
population, $y_i$ is the fraction of allele $i$ in the second
population. A second sum is over the loci $l$ and $a_m$ is the
total number of alleles, but if we work in the gamete space,
then $x_i$ and $y_i$ refer to gamete frequencies and $l=1$.

$NEI$ is defined as:
\begin{equation}\label{nei}
NEI\equiv -\log\left(\frac{\sum_l \sum_i x_iy_i}{\sqrt{\sum_l \sum_i x_i^2\sum_l \sum_i y_i^2}}\right).
\end{equation}

The main statistical difference between $NEI$ and $CS$ with
$JSD$, as we are going to show, is that $JSD$ incorporates a
weighting system for the different population sizes, while
$NEI$ and $CS$ don't.

\begin{figure}[!ht]\center
\includegraphics[width=.7\textwidth]{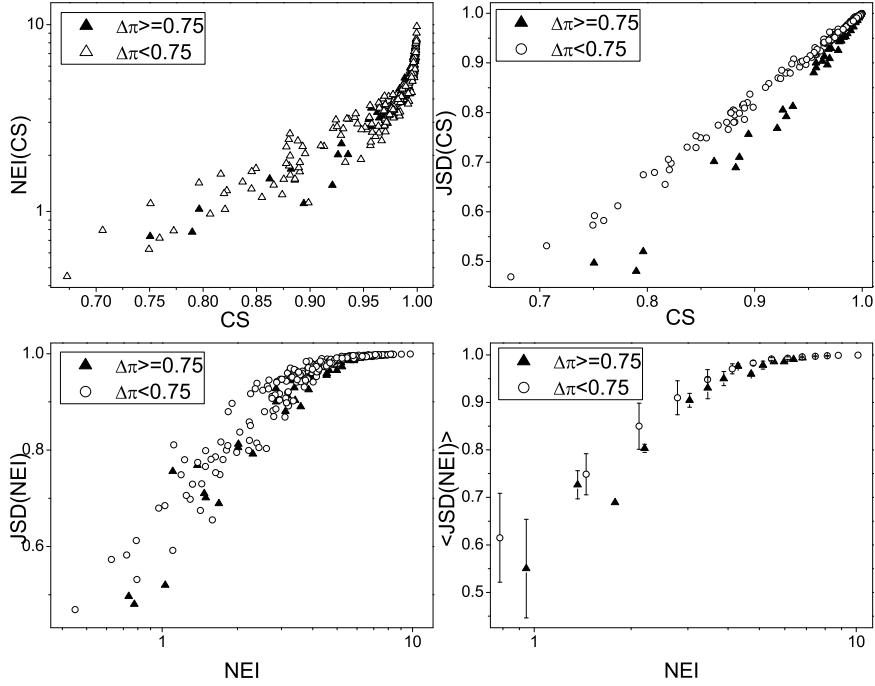}
\caption{Measures performed for the CN dataset.
      Top-left: correlations between $NEI$ and $CS$. Top-right: $JSD$ versus $CS$.
Bottom-left: $JSD$ versus $NEI$. Bottom-right: $< JSD(NEI) >$ versus $ NEI$.}
\end{figure}

In Fig.1 we show the correlations between those
measures, as measured in the CN dataset. For a reason that will
be clear below we define the parameter $\Delta
\pi=|\pi_1-\pi_2|$ as the absolute value of the difference of
the statistical weight between  population 1 and 2. We
represent with black triangles the distances between
populations whose weight difference $\Delta \pi$ is larger or
equal than 0.75 and with white circles the distances between
those populations whose weight difference $\Delta \pi$ is less
than 0.75.

In  the top-left panel we show the plot for $NEI$ versus $CS$.
The main difference between $NEI$ and $CS$ is that $NEI$
considers the variance of the gamete distributions, while $CS$
doesn't. We see that the measures are well correlated with the
differences given by the variances of the populations.

In the top-right panel we show the plot of $JSD$ versus $CS$.
Also in this case the measures are positively correlated, even
if two different branches appear in the plot. Those two
branches appear to be well represented by the two different
categories for $\Delta \pi$. In fact when the population sizes
are very different, $\Delta \pi\geq 0.75$, we see that $CS$
overestimates the distance between them in respect to $JSD$.

In the bottom-left panel we show the plot of $JSD$ versus
$NEI$. Again we can see that the measures are well correlated,
but also in this case $NEI$ overestimates the distance between
two populations when $\Delta \pi\geq 0.75$, even if this is
less evident that in the previous case. Nevertheless if we look
at the plot of the average value of the $JSD$ corresponding to
the same $NEI$, $<JSD(NEI)>$, on the bottom-right panel, we can
see that it is true in average, since each black triangle is
below the correspondent white circle.

 The \textbf{average square distance} $ASD$ between population $A$ and $B$ is defined as \cite{3}:
\begin{equation}\label{3}
ASD_k\equiv \sum_{i,j}(i-j)^2f_if_j = (\mu_A^k-\mu_B^k)^2+V_A^k+V_B^k =(\delta \mu^k)^2+V_A^k+V_B^k,
\end{equation}
where $i$, $j$ are the repetition numbers of allele $i\epsilon
A$ and $j \epsilon B$ and $f_{i,j}$ its frequency at locus $k$.
Then $\mu_X^k=\sum i\cdot f_i$ is the average repetition number
at locus $k$ for population $X$ and $V_{X}^k=\sum
f_{i}(i-\mu_{X}^k)$ is the variance in the repetition number of
population $X$ at locus $k$. Eq.\ref{3} has to be averaged on
the loci then.
\begin{equation}
ASD=\frac{\sum_{l}^nASD_k}{l}.
\end{equation}

 The \textbf{Goldstein distance}  $GD$ between two populations $A$ and $B$
 defined by a set of alleles is defined at the locus $k$ as \cite{4}:
\begin{equation}\label{1}
(\delta \mu^k)^2\equiv (\mu_A^k-\mu_B^k)^2,
\end{equation}
where $\mu_X^k$ is the average number of repetitions for population $X$ at locus $k$.

In the case of $n$ different loci Eq.\ref{1} is averaged on the different loci:
\begin{equation}\label{4}
GD\equiv \frac{\sum_{k=1}^n (\delta \mu^k)^2}{n}.
\end{equation}

 $GD$ was introduced as an improvement of $ASD$  \emph{"because distances based
on the infinite-alleles model are nearly linear with time
immediately following isolation, it is not worthwhile to use
$ASD$ with very closely related groups"} \cite{4}.

The difference between $GD$ and $ASD$, as it is clear from
Eq.\ref{3} and Eq.\ref{4}, resides on the presence in $ASD$ of
the variance of the attributes.

To understand the difference between $JSD$ and $GD$ in the
gamete space, we have to keep in mind  the representation of
the genet in the gamete space as explained in the main paper.
The gamete space is a $n$-dimensional space, each dimension
representing a locus, where a diploid genet is represented by a
set of $2^n$ points. For a population $X$ such a distribution
of points has a centre of mass, whose coordinates $\mu_X^k$
($k=1,..,n$) are given by the average repetition number for
each locus $\mu_X^k=\sum i\cdot f_i$. Hence, given two
populations $A$ and $B$, the distance between their average
point in the attribute space is given by $\sqrt{\sum_{i=k}^n
(\mu_A^k-\mu_B^k)^2}=\sqrt{n\cdot GD}$ (see Eq.\ref{4}).

\begin{figure}[!ht]\center
\includegraphics[width=.7\textwidth]{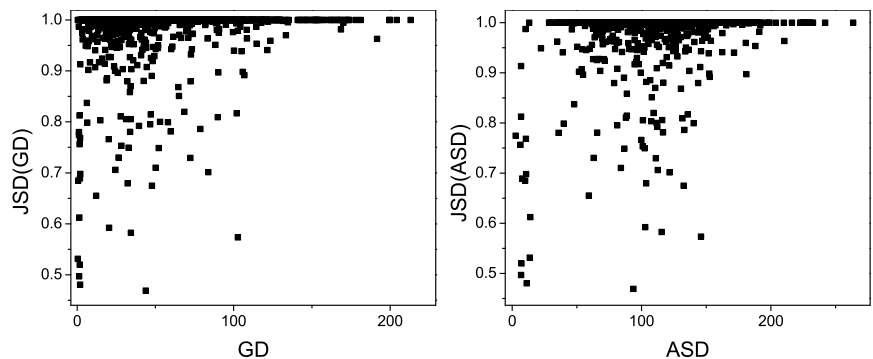}
\caption{Measures performed for the CN dataset.
      Left panel: $JSD$ versus $GD$. Right panel: $JSD$ versus $ASD$}
\end{figure}

Hence $GD$ is the square root of the distance between the
centre of mass of the populations as represented in the gamete
space. $JSD$ instead is a punctual information measure that
considers all the point correlations between the two
populations. Then $JSD$ and $GD$ could be correlated, this does
not occur always. For instance there is an extreme case where
population $A$ and $B$ have the same centre of mass in the
gamete space, so that $GD=0$, but not a single common gamete so
that $JSD=1$. We can observe this in Fig.2, where we
show evidence for not very strong correlations between $JSD$
and $GD$ in the left panel and between $JSD$ and $ASD$ in the
right panel, as measured for the CN dataset.


\end{bmcformat}


\end{document}